\documentclass{JHEP3}
\usepackage{graphicx}
\usepackage{amsmath,epsfig}
\usepackage{amssymb,amsfonts}
\usepackage{latexsym}
\usepackage{epsfig}

\relax

\def\be{\begin{equation}}
\def\ee{\end{equation}}

\newcommand{\rc}{\nonumber\\}

\newcommand{\bear}{\begin{eqnarray}}
\newcommand{\bea}{\begin{eqnarray}}
\newcommand{\eear}{\end{eqnarray}}
\newcommand{\eea}{\end{eqnarray}}

\newbox\pippobox

\def\II{\relax{\rm I\kern-.18em I}}

\def\sp{\;\;\;,\;\;\;}

\title{An AdS/QCD model from Sen's tachyon action}

\author{Ioannis Iatrakis$^a$, Elias Kiritsis$^{a}$\footnote{On leave of absence from APC,
Universit\'e Paris 7, (UMR du CNRS 7164).},
\'Angel Paredes$^b$\\
~\\
$^a$ Crete Center for Theoretical Physics,
Department of Physics, University of Crete, 71003 Heraklion, Greece\\
~\\
$^b$  Departament de F\'\i sica Fonamental and ICCUB Institut de
Ci\`encies del Cosmos, Universitat de Barcelona, Mart\'\i\ i Franqu\`es, 1,
E-08028, Barcelona, Spain.}

\preprint{CCTP-2010-2~~~~~~~~~~~~~~\\}

\abstract{We construct a new, simple phenomenological model along the lines of AdS/QCD.
The essential new ingredient is the brane-antibrane effective action including the open string tachyon
 proposed by Sen. Chiral symmetry breaking happens because of tachyon dynamics.
 We fit a large number of low-spin meson masses at the 10\%-15$\%$ level. 
The only free parameters involved in the fits correspond to the overall QCD-scale and the
quark masses. Several aspects of previous models are qualitatively improved.}


\begin{document}

\section{Introduction}
\setcounter{equation}{0}

Understanding the strong dynamics underlying many observations related to the strong
interaction remains an unsolved problem. Much progress has been made to date with different
methods but new insights  are always
welcome. A recent development that has led to reconsidering the strong interaction has been
the AdS/CFT duality. This has been applied to obtain new insights on QCD phenomenology.
In this paper, we focus on the meson spectrum - (see \cite{Erdmenger:2007cm}
for a review of the gauge-gravity literature on the issue).

There are two main ways to address the problem at hand. Top-down approaches
use string theory from first principles in order to build dual theories as close as possible to QCD. 
Notable examples are \cite{topdown},
\cite{ss}.
On the other hand, bottom-up approaches, starting from the works \cite{Polchinski:2001tt,Erlich:2005qh,Pomarol1},
use known QCD features to develop holographic models that are only inspired by string theory.
The model we present here is more of the second kind, but includes the main stringy ingredients we expect from first principles, namely
the effective action controlling the chiral dynamics.
Our main observation is that  merging this stringy input  of top-down holographic models for flavor
 allows us to improve the existing
bottom-up models both  qualitatively and quantitatively.

\section{The model}

\subsubsection*{Background and action}

In \cite{ckp}, it was shown, quite generally, that effective actions
for brane-antibrane systems derived from
string theory \cite{Sen:2003tm} encode a set of qualitative features related to chiral symmetry breaking and
QCD at strong coupling.
The goal of this paper is to build a concrete model within the framework of
\cite{ckp}. We will consider the simplest smooth gravitational background that is asymptotically AdS,
 while having a confining IR in the same spirit as \cite{Witten:1998zw}. This turns out to be the $AdS_6$ soliton, which was shown to
be a solution to the two-derivative approximation of subcritical string theory and used as a toy
model for certain aspects of 4d Yang-Mills in
\cite{Kuperstein:2004yf}. The metric reads:
\be
ds_6^2 = \frac{R^2}{z^2} \left[ dx_{1,3}^2 +  f_\Lambda^{-1} dz^2 + f_\Lambda\, d\eta^2 \right]
\label{adsbh6}
\ee
with
$f_\Lambda=1-\frac{z^5}{z_\Lambda^5}$.
The coordinate $\eta$ is periodically identified and $z\in [0,z_\Lambda]$.
The dilaton is constant and we do not write the RR-forms since they do not play any role in
the following. We now consider a D4-$\bar {\rm D}4$  pair, located at fixed $\eta$ in this
 background\footnote{In five-dimensional holographic models of QCD, the flavor branes are expected to be a
  D4-$\bar {\rm D}4$ system, \cite{ihqcd}.}. We
write the action proposed by Sen \cite{Sen:2003tm} as
\be
S= - \int d^4x dz  V(|T|)
\left(\sqrt{-\det {\bf A}_L}+\sqrt{-\det {\bf A}_R}\right)
\label{generalact}
\ee
 The objects inside the
square roots are defined as:
\be
{\bf A}^{(i)}_{MN}=g_{MN} + \pi \alpha' \left[2F^{(i)}_{MN}
+   \left((D_M T)^* (D_N T)+(M\leftrightarrow N)\right)\right]
\label{Senaction}
\ee
where $M,N=1,\dots,5$, the field strengths $F^{(i)}_{MN}=
\partial_M A^{(i)}_N - \partial_N A^{(i)}_M$ and the covariant derivative of the
tachyon is
$D_M T = (\partial_M + i A_M^L- i A_M^R)T$.
The active fields in (\ref{generalact}), (\ref{Senaction}) are two 5-d gauge fields
and a complex scalar $T=\tau\, e^{i\theta}$,
which are dual to the low-lying quark bilinear operators
that correspond to states with $J^{PC}=1^{--},1^{++},0^{-+},0^{++}$ -;
see \cite{ckp} for details.
In the action of \cite{Sen:2003tm}, the transverse scalars (namely $\eta$ in the
present case) are also present. We have discarded  them when writing
(\ref{Senaction})
since they do not have any
interpretation in terms of QCD fields.
Accordingly, even if the background (\ref{adsbh6}) is six-dimensional,
the holographic model for the hadrons is effectively five-dimensional and, in
fact, its field content coincides with those of \cite{Erlich:2005qh},
\cite{Pomarol1}.
For the tachyon potential we take, as the simplest possibility, the
one computed in boundary string field theory for an unstable D$p$-brane
in flat space
\cite{Kutasov:2000aq}, although one should keep in mind that this
expression for $V$ is not
top-down derived for the present situation.
In the present conventions,
$V= {\cal K}\, e^{-\frac{\pi}{2} \tau^2}$,
where ${\cal K}$ is an overall constant that will play no role in the following
since it does not enter the meson spectrum computation (it is important though, in the
normalization of correlators when computing for
instance decay constants \cite{Erlich:2005qh}, \cite{Pomarol1}, \cite{long}).
The tachyon mass is $m_T^2 = -\frac{1}{2\alpha'}$ and 
 we will impose:
$R^2= 6 \alpha'$ in order to have $m_T^2 R^2= - 3$. This should not be interpreted as a modification
of the background due to the branes, but just as a (bottom-up) choice of the string scale that
controls the excitations of those branes,
such that the bifundamental scalar $T$ is dual to an operator of dimension 3,
as in \cite{Erlich:2005qh},
\cite{Pomarol1}. 
Since the AdS radius is not parametrically larger than $\alpha'$, the two-derivative action cannot
be a controlled low enegy approximation to string theory. This is the main reason why a 
model of this kind cannot be considered of top-down nature.
Notice the value of $R^2$ we take differs from the one used in
\cite{Kuperstein:2004yf}.

\subsubsection*{The tachyon vacuum and chiral symmetry breaking}

As shown in \cite{ckp}, an essential ingredient of the present  framework is that the generation of the correct flavor anomaly on the flavor branes
 requires the tachyon modulus $\tau$ to diverge somewhere. Therefore, $\tau$ must have a nontrivial vev which breaks the
chiral symmetry. From the action (\ref{generalact}) we obtain the equation
determining $\tau(z)$:
\be
\tau'' - \frac{4 \pi\, z\, f_\Lambda}{3} \tau'^3
+ ( - \frac{3}{z}  + \frac{f_\Lambda'}{2f_\Lambda})\tau'
+ \left(\frac{3}{  z^2 f_\Lambda} +\pi\, \tau'^2 \right) \tau =0
\label{taueq}
\ee
where the prime stands for derivative with respect to $z$.
Near $z=0$, the solution can be expanded in terms of two integration
constants as:
\be
\tau = c_1 z + \frac{\pi}{6}c_1^3 z^3 \log z + c_3 z^3 + {\cal O}(z^5)
\label{tauUVexpan}
\ee
where, on general AdS/CFT grounds, $c_1$ and $c_3$ are related to the quark mass and condensate
(see \cite{long} for a careful treatment).
From  (\ref{taueq}), we find that $\tau$ can diverge only at $z=z_\Lambda$.
There is a one-parameter family of diverging solutions in the IR:
\be
\tau = \frac{C}{(z_\Lambda - z)^{\frac{3}{20}}}- \frac{13}{6 \pi C}(z_\Lambda - z)^{\frac{3}{20}}
+ \dots
\label{IRdiver}
\ee
The interpretation is the following: for a given $c_1$ (namely quark
mass\footnote{For the present work, we will just use that $c_1$ is proportional to $m_q$.
Finding the proportionality coefficient requires normalizing the action and fields as
in \cite{DaRold:2005vr}.}) fixed in the UV
(near $z=0$),
the value of $c_3$ (namely the quark condensate) is determined dynamically by requiring that
the numerical integration of (\ref{taueq}) leads to the physical IR (near $z=z_\Lambda$)
behavior  (\ref{IRdiver}).
Hence, for any value\footnote{In practice, we have been able to perform numerics in a controlled
manner only for $0 \leq c_1 < 1$.}
of $c_1$, one can obtain numerically the function for the vev $\langle \tau \rangle$.

\subsubsection*{Meson spectrum: Numerical results}

There is a rather standard method for computing the meson spectrum in
holographic models, see \cite{Erdmenger:2007cm} for a review.
Each bulk field is dual to a boundary operator and its linearized perturbation
can be obtained after expanding (\ref{generalact}). By looking at normalizable
fluctuations of the bulk fields, one typically encounters discrete towers of masses
for the  physical states with the corresponding quantum numbers. 
Thus, for a fixed value of $c_1$, we have a Sturm-Liouville problem for each bulk mode.
This can be solved numerically, using a standard shooting technique.
By computing the different towers at different values of $c_1$, we found the
following expressions to be very good approximations to the numerical
results, in the range $0< c_1<1$ where we could perform the numerics reliably.
For the vectors:
\bear
z_\Lambda\,m_V^{(1)} &=& 1.45 + 0.718 c_1 \sp
z_\Lambda\,m_V^{(2)} = 2.64 + 0.594 c_1 \rc
z_\Lambda\,m_V^{(3)} & =& 3.45 + 0.581 c_1 \sp
 z_\Lambda\,m_V^{(4)} = 4.13 + 0.578 c_1 \rc
 z_\Lambda\,m_V^{(5)} &=& 4.72 + 0.577 c_1 \sp
z_\Lambda\,m_V^{(6)} = 5.25 + 0.576 c_1 .
\label{vectormassesfit}
\eear
For the axial vectors:
\bear
z_\Lambda\,m_A^{(1)} &=& 1.93 + 1.23 c_1\sp 
z_\Lambda\,m_A^{(2)} = 3.28 + 1.04 c_1 \rc
 z_\Lambda m_A^{(3)} &=& 4.29 + 0.997 c_1 \sp
 z_\Lambda m_A^{(4)} = 5.13 + 0.975 c_1 \rc
z_\Lambda\,m_A^{(5)} &=& 5.88 + 0.962 c_1 \sp
 z_\Lambda\,m_A^{(6)} = 6.55 + 0.954 c_1 .
\eear
For the pseudoscalars:
\bear
z_\Lambda\,m_P^{(1)} &=& \sqrt{2.47 c_1^2 + 5.32 c_1} \sp
z_\Lambda\,m_P^{(2)} = 2.79 + 1.16 c_1 \rc
z_\Lambda\,m_P^{(3)} &=& 3.87+ 1.08 c_1 \sp
 z_\Lambda\,m_P^{(4)} = 4.77 + 1.04 c_1 \rc
 z_\Lambda\,m_P^{(5)} &=& 5.54 + 1.01 c_1 \sp
 z_\Lambda\,m_P^{(6)} =6.24 + 0.997 c_1.
\label{psmassesfit}
\eear
For the scalars:
\bear
z_\Lambda\,m_S^{(1)} &=& 2.47 + 0.683 c_1 \sp
z_\Lambda\,m_S^{(2)} = 3.73 + 0.488 c_1 \rc
z_\Lambda\,m_S^{(3)} &=& 4.41 + 0.507 c_1\sp
z_\Lambda\,m_S^{(4)} = 4.99 + 0.519 c_1\rc
z_\Lambda\,m_S^{(5)} &=& 5.50 + 0.536 c_1\sp
z_\Lambda\,m_S^{(6)} = 5.98 + 0.543 c_1.
\label{scalarmassesfit}
\eear 
It turns out  that meson masses increase linearly with $c_1$. Namely, they
increase linearly with the bare quark mass, as expected from an
expansion in $m_q$ and in qualitative agreement with lattice
results, see for instance \cite{Laermann:2001vg},\cite{DelDebbio:2007wk},\cite{Bali:2008an}.
The exception, of course, is the first pseudoscalar for which
$m_\pi$ is proportional to $\sqrt{m_q}$ (for small $m_q$), as expected from
the Gell-Mann-Oakes-Renner relation. Actually,
the behaviour  $m_\pi=\sqrt{b\, m_q + d\, m_q^2}$  was
also found in the lattice \cite{Laermann:2001vg}.

\section{Fitting the meson spectrum}

We now proceed to make a phenomenological comparison of the results
of (\ref{vectormassesfit})-(\ref{scalarmassesfit}) to the
experimental values quoted by the Particle Data Group \cite{Amsler:2008zzb}.
Obviously, we can only model those mesons with
$J^{PC}=1^{--},1^{++},0^{-+},0^{++}$.
From \cite{Amsler:2008zzb}, we will just take the central value quoted for each
resonance. We do not discuss decay widths here (in the strict $N_c \to \infty$ limit
they are of course zero).

\subsection*{Isospin 1 mesons}

We start by looking at mesons composed of the light quarks $u$ and
$d$. In particular, we discuss the isovectors. In table
\ref{table1}, we show all the
mesons listed in the meson summary table of
\cite{Amsler:2008zzb} under {\it light unflavored mesons}
which have
isospin 1
and the $J^{PC}$\,'s present in our model.
The only exception is $a_0(980)$, which is considered to be a four-quark state
\cite{Amsler:2008zzb}.
We have fitted the  parameters of the model to these observables
by minimizing the rms error $\varepsilon_{rms}=100\times \frac{1}{\sqrt n}\left(\sum_O
\left(\frac{\delta O}{O}\right)^2\right)^{\frac12}$, where $n=8-2=6$ is the
number of observables minus the number of parameters. We obtain for the parameters
\be
z_\Lambda^{-1}= 522 \rm{MeV} \,\,,\qquad\quad
c_{1,l} = 0.0125
\label{fitzlambda}
\ee
with $\varepsilon_{rms}=12\%$.
\begin{table}[h]
\begin{center}
\begin{tabular}{|c|c|c|c|}
\hline
$J^{PC}$ & Meson & Measured (MeV) & Model (MeV)\\
\hline
$1^{--}$ & $\rho(770)$  & 775 &  762  \\
\cline{2-4}
 & $\rho(1450)$  & 1465 &  1379  \\
\cline{2-4}
& $\rho(1700)$  & 1720 &  1806  \\
\cline{1-4}
$1^{++}$ & $a_1 (1260)$  & 1230 &  1015  \\
\cline{1-4}
$0^{-+}$ & $\pi_0$  & 135.0 &  135.1  \\
\cline{2-4}
 & $\pi(1300)$  & 1300 &  1462  \\
\cline{2-4}
& $\pi(1800)$  & 1816 &  2026 \\
\cline{1-4}
$0^{++}$ & $a_0(1450)$  & 1474 &  1295  \\
\hline
\end{tabular}
\end{center}
\caption{A comparison of the results of the model to the experimental values for light
unflavored meson masses.}
\label{table1}
\end{table}

In table \ref{table2}, we display the resonance masses with isospin 1 listed in
\cite{Amsler:2008zzb} under {\it other light unflavored mesons}. These are namely states
considered as ``poorly established that thus require confirmation". For the
results given by our model, we use (\ref{fitzlambda}) and therefore no further parameter
is fitted here. For this set of observables we get $\varepsilon_{rms}=24\%$, where
we have inserted $n=11-0=8$. One should keep in mind that it is plausible that some of
these ``unconfirmed" states may not be real or may be misinterpreted as part of the 
meson towers. In this sense, our model seems to favor the $\rho(2150)$ as the fourth member
of the $\rho$-meson tower \footnote{Indeed, there is much more experimental evidence for $\rho(2150)$ than
for $\rho(1900)$ or $\rho(1570)$. We thank S. Eydelman for very useful explanations.}.
We have not included $\rho(1570)$ in  table \ref{table2}
 because its excitation number is
smaller than $\rho(1700)$, which was included in the previous fit.
In case $\rho(1570)$ gets confirmed as a member of this tower, the fit should be redone.
We observe that the model
tends to consistently overestimate the masses of the excited axial vectors and pions. 
This is connected to
the fact the model yields a Regge slope for axial mesons larger than the one for the vectors
\cite{ckp}. If the experimental results of table \ref{table2} are confirmed, one should
think of improving the model in order to ameliorate this discrepancy.
\begin{table}[h]
\begin{center}
\begin{tabular}{|c|c|c|c|}
\hline
$J^{PC}$ & Meson & Measured (MeV) & Model (MeV)  \\
\hline
$1^{--}$ & $\rho(1900)$  & 1900 &  2159  \\
\cline{2-4}
 & $\rho(2150)$  & 2150 &  2467  \\
\cline{2-4}
 & $\rho(2270)$  & 2270 &  2746  \\
\cline{1-4}
$1^{++}$ & $a_1 (1640)$  & 1647 &  1721  \\
\cline{2-4}
 & $a_1 (1930)$  & 1930 &  2245  \\
\cline{2-4}
 & $a_1(2096)$  & 2096 &  2686  \\
 \cline{2-4}
 & $a_1(2270)$  & 2270 &  3073  \\
 \cline{2-4}
 & $a_1(2340)$  & 2340 &  3423  \\
\cline{1-4}
$0^{-+}$ & $\pi(2070)$  & 2070 &  2493  \\
\cline{2-4}
 & $\pi(2360)$  & 2360 &  2899  \\
\cline{1-4}
$0^{++}$ & $a_0(2020)$  & 2025 &  1952 \\
\hline
\end{tabular}
\end{center}
\caption{A comparison of the results of the model to the experimental values for
other light
unflavored meson masses.}
\label{table2}
\end{table}

\subsection*{ $s \bar s$ states}

A  nice feature of the present model is that it incorporates the
dependence of the hadron masses on the quark mass. This allows us to
study $s \bar s$ states.
More precisely, it allows us to discuss
``hypothetical states" with quark content $s \bar s$ assuming no mixing
with other states. In the real world, the mixing for pseudoscalars and scalars is
important (see chapter 14 of \cite{Amsler:2008zzb}), and therefore it is not
possible to compare directly the outcome of the model to the experimental results.
Nevertheless, as in \cite{Allton:1996yv}, we can estimate the masses of these
``hypothetical" $s \bar s$ mesons from the light-strange and light-light mesons.
Then, using quotation marks for the hypothetical states,
and using the quark model classification (table 14.2 of \cite{Amsler:2008zzb})
we can write
$m(``\eta")= \sqrt{ 2 m_K^2 - m_\pi^2}$, $m(``\phi(1020)") = 2 m(K^*(892))-
m(\rho(770))$, $m(``\eta(1475)")=2m(K(1460))-m(\pi(1300))$, etc.
Keeping the value of $z_\Lambda$ found in (\ref{fitzlambda}),
we fit the value of $c_1$ associated with the strange quark to the
``experimental" values of table \ref{table3}, obtaining.
\be
c_{1,s}=0.317
\label{fitc1s}
\ee
The rms error for this set of observables ($n=6-1$) is
$\varepsilon_{rms}=10\%$.
\begin{table}[h]
\begin{center}
\begin{tabular}{|c|c|c|c|}
\hline
$J^{PC}$ & Meson & Measured (MeV) & Model (MeV)  \\
\hline
$1^{--}$ & $``\phi(1020)"$  & 1009 &  876  \\
\cline{2-4}
 & $``\phi(1680)"$  & 1363 &  1474  \\
\cline{1-4}
$1^{++}$ & $``f_1 (1420)"$  & 1440 &  1210  \\
\cline{1-4}
$0^{-+}$ & $``\eta"$  & 691 &  725  \\
\cline{2-4}
 & $``\eta(1475)"$  & 1620 &  1647  \\
\cline{1-4}
$0^{++}$ & $``f_0(1710)"$  & 1386 &  1403  \\
\hline
\end{tabular}
\end{center}
\caption{A comparison of the results of the model to the
hypothetical states  with $s \bar s$ quark content.
}
\label{table3}
\end{table}

\subsection*{Further comments}

We end this section by commenting on the estimates given by the model
on two other physical quantities. Without fixing the proportionality
coefficient between $c_1$ and the quark mass, from (\ref{fitzlambda}), (\ref{fitc1s}),
we can infer the ratio of the strange quark  to the light quark mass:
$\frac{2 m_s}{m_u + m_d} \approx \frac{c_{1,s}}{c_{1,l}} \approx 25$.
Moreover, the background studied here experiences a first order deconfinement phase
transition in full analogy with \cite{Witten:1998zw}.
The deconfinement temperature is given by
$T_{deconf} = \frac{5}{4\pi\, z_\Lambda} \approx 208$ MeV.
Both of these values are close to the experimental values.

\section{Conclusions}

We have built a new phenomenological model for the meson sector of QCD.
In this paper we have discussed the mass spectrum. We note the simplicity of the construction,
whose essential point is the use of Sen's action \cite{Sen:2003tm}
including the open string tachyon field. We have applied it to one of the  simplest backgrounds
exhibiting  confinement \cite{Kuperstein:2004yf}.
Despite the minimal  input, we have found the following interesting qualitative properties:

\begin{itemize}

\item The model includes towers of excitations with $J^{PC}=1^{--},1^{++},0^{-+},0^{++}$,
namely all low-lying operators that do not need a dual excited stringy state.

\item Chiral symmetry breaking is consistently realized. Moreover, the value of the quark
condensate is computed dynamically and is not a tunable input. Hence, the number of tunable
parameters coincides with those present in QCD: they are just the dynamically generated scale
and the quark masses.

\item We find Regge trajectories for the excited states $m_n^2\, \propto \, n$, as in the
soft wall model \cite{Karch:2006pv}. This allows good predictions for the higher
excitations, as opposed to the hard wall model \cite{Erlich:2005qh,Pomarol1}. Notwithstanding,
the Regge slope for axial vectors is bigger than the one for vectors. This fact requires further
study.

\item Our model incorporates the increase of the vector meson masses due to the increase of
quark masses, as $m_\rho \approx k_1 + k_2\, m_\pi^2$ for small $m_\pi$.
\end{itemize}

Previous AdS/QCD models present some of these properties, but as far as we know, no existing
model is able to encompass all of them, see \cite{Sui:2009xe} for recent related
discussions. We briefly comment on the three benchmark models:
the Sakai-Sugimoto model \cite{ss} misses the first and third points listed above,
the hard wall model  \cite{Erlich:2005qh,Pomarol1} misses the third one and
partially the second one; and the
soft wall model \cite{Karch:2006pv} misses the second one. All
of these models \cite{ss,Erlich:2005qh,Pomarol1,Karch:2006pv} and variations
thereof
fail to get the fourth point (although it is worth mentioning that
D3D7 models with abelian flavor symmetry do capture the physics of this fourth
point, see section 6.2.3 of \cite{Erdmenger:2007cm}).

Moreover, the quantitative matching shown in tables \ref{table1} and \ref{table3} with
the central values of the meson resonances is excellent, at the 10\%-15\% level.
This is a typical accuracy of AdS-QCD-like models (a recent example is
\cite{terBrod}, which accounts for excited spin states of the $\rho$ and $\omega$ families).
Since the systematic error produced by quenching is of the order of 10\%
\cite{Aoki:1999yr} and the
differences between quenched lattice computations with $N_c=3$ and $N_c = \infty$
are again of the order of 10\% \cite{DelDebbio:2007wk,Bali:2008an},
it would be unexpected to get a better accuracy from
any model of the kind presented here.

It would be of utmost interest to generalize
the set-up to the non-abelian case, allowing several quark flavors, but
this is beyond the scope of the present work.

\vskip .2cm
\centerline{\bf Acknowledgments}

\noindent
This work was  partially supported by  a European Union grant FP7-REGPOT-2008-1-CreteHEP
 Cosmo-228644.
The research of A.P is
supported by grants FPA2007-66665C02-02 and DURSI
2009 SGR 168, and by the CPAN CSD2007-00042 project of the
Consolider-Ingenio 2010 program.
I.I. has been supported by Manasaki Graduate Scholarship and the Physics department of the University of Crete until September 2009 and by A.S. Onassis Foundation Graduate Scholarship from October 2009.



\begin{thebibliography}{99}


\bibitem{Erdmenger:2007cm}
  J.~Erdmenger, N.~Evans, I.~Kirsch and E.~Threlfall,
  Eur.\ Phys.\ J.\  A {\bf 35}, 81 (2008).

\bibitem{topdown}
I.~R.~Klebanov and M.~J.~Strassler,
  JHEP {\bf 0008}, 052 (2000). J.~Babington, J.~Erdmenger, N.~J.~Evans, Z.~Guralnik and I.~Kirsch,
  Phys.\ Rev.\  D {\bf 69}, 066007 (2004).
 M.~Kruczenski, D.~Mateos, R.~C.~Myers and D.~J.~Winters,
  JHEP {\bf 0405}, 041 (2004).


\bibitem{ss}
  T.~Sakai and S.~Sugimoto,
  Prog.\ Theor.\ Phys.\  {\bf 113} (2005) 843.


\bibitem{Polchinski:2001tt}
  J.~Polchinski and M.~J.~Strassler,
  Phys.\ Rev.\ Lett.\  {\bf 88}, 031601 (2002)
  [arXiv:hep-th/0109174].



\bibitem{Erlich:2005qh}
  J.~Erlich, E.~Katz, D.~T.~Son and M.~A.~Stephanov,
  Phys.\ Rev.\ Lett.\  {\bf 95}, 261602 (2005).

\bibitem{Pomarol1}
L.~Da Rold and A.~Pomarol,
  Nucl.\ Phys.\  B {\bf 721}, 79 (2005).


\bibitem{ckp}
  R.~Casero, E.~Kiritsis and A.~Paredes,
  Nucl.\ Phys.\  B {\bf 787}, 98 (2007).


\bibitem{Sen:2003tm}
  A.~Sen,
  Phys.\ Rev.\  D {\bf 68}, 066008 (2003).

\bibitem{Witten:1998zw}
  E.~Witten,
  Adv.\ Theor.\ Math.\ Phys.\  {\bf 2}, 505 (1998).


\bibitem{Kuperstein:2004yf}
  S.~Kuperstein and J.~Sonnenschein,
  JHEP {\bf 0411}, 026 (2004).

  \bibitem{ihqcd}
  F.~Bigazzi, R.~Casero, A.~L.~Cotrone, E.~Kiritsis and A.~Paredes,
  JHEP {\bf 0510}, 012 (2005);
  U.~Gursoy and E.~Kiritsis,
  JHEP {\bf 0802} (2008) 032;
   U.~Gursoy, E.~Kiritsis and F.~Nitti,
  JHEP {\bf 0802} (2008) 019;
 U.~Gursoy, E.~Kiritsis, L.~Mazzanti and F.~Nitti,
  JHEP {\bf 0905} (2009) 033.

\bibitem{Kutasov:2000aq}
 D.~Kutasov, M.~Marino and G.~W.~Moore,
  JHEP {\bf 0010} (2000) 045.





\bibitem{long}
  I.~Iatrakis, E.~Kiritsis and A.~Paredes,
{\it in preparation}








\bibitem{DaRold:2005vr}
  L.~Da Rold and A.~Pomarol,
  JHEP {\bf 0601}, 157 (2006).


\bibitem{Laermann:2001vg}
  E.~Laermann and P.~Schmidt,
  Eur.\ Phys.\ J.\  C {\bf 20}, 541 (2001).

\bibitem{DelDebbio:2007wk}
  L.~Del Debbio, B.~Lucini, A.~Patella and C.~Pica,
  JHEP {\bf 0803}, 062 (2008).

\bibitem{Bali:2008an}
  G.~S.~Bali and F.~Bursa,
  JHEP {\bf 0809}, 110 (2008).

\bibitem{Amsler:2008zzb}
  C.~Amsler {\it et al.}  [Particle Data Group],
  Phys.\ Lett.\  B {\bf 667}, 1 (2008).

\bibitem{Allton:1996yv}
  C.~R.~Allton, V.~Gimenez, L.~Giusti and F.~Rapuano,
  Nucl.\ Phys.\  B {\bf 489}, 427 (1997).


\bibitem{Karch:2006pv}
  A.~Karch, E.~Katz, D.~T.~Son and M.~A.~Stephanov,
  Phys.\ Rev.\  D {\bf 74}, 015005 (2006).


\bibitem{Sui:2009xe}
T.~Gherghetta, J.~I.~Kapusta and T.~M.~Kelley,
  Phys.\ Rev.\  D {\bf 79}, 076003 (2009);
  Y.~Q.~Sui, Y.~L.~Wu, Z.~F.~Xie and Y.~B.~Yang,
  Phys.\ Rev.\  D {\bf 81}, 014024 (2010).




\bibitem{terBrod}
G.~F.~de Teramond and S.~J.~Brodsky,
  arXiv:1001.5193 [hep-ph].


\bibitem{Aoki:1999yr}
  S.~Aoki {\it et al.}  [CP-PACS Collaboration],
  Phys.\ Rev.\ Lett.\  {\bf 84}, 238 (2000).






\end{thebibliography}
\end{document}